\documentclass[11pt,twoside]{article}

\usepackage{asp2006_ecl}
\usepackage{epsf}
\usepackage{psfig}
\usepackage{lscape}
\usepackage{epsfig}

\begin{document}
\title{Panel discussion II: Reconciling observations and modeling of
  star formation at high redshifts}   
\author{Johan H. Knapen (Reporter)}
\affil{Instituto de Astrof\'\i sica de Canarias, E-38200 La Laguna, Spain}

\begin{abstract} 

This is the written account of the second of two panel discussions, on
{\it Reconciling observations and modeling of star formation at high
redshifts}. The chair of the panel was Pavel Kroupa, and panel members
were Marc Balcells, John Beckman, Christopher Conselice, and Joseph
Silk. After a short introduction by each of the panelists, panel and
audience entered into a lively discussion, centered around the
following six themes: the mass function of pre-stellar gas clouds; a
possible top-heavy initial mass function at high redshifts versus
mini-quasars as the first sources of ionization; the integrated
galactic initial mass function; possible differences in specific star
formation rates in disks and in massive galaxies; whether merging
rates yield a wrong prediction for massive galaxies, and what is the
physics behind the onset of the red sequence of galaxies; and the case
of dark matter-dominated dwarf galaxies versus tidal dwarf galaxies.

\end{abstract}


\medskip

{\sc Kroupa:} Welcome to this second panel discussion, focused on
observations and modeling of star formation (SF) in the high redshift
Universe. Each of our panelists will start by giving a brief
introduction to an aspect of this general topic that they find
particularly interesting, and which they propose for further
discussion. Let me ask John Beckman to start.

\medskip

{\sc Beckman:} The point I would like to make is not (at any rate not
yet) on SF at high redshifts but on the initial mass function
(IMF). The question is whether there is a universal IMF wherever and
whenever stars have formed. It would imply, for example, that
metallicities in stellar populations could be derived more precisely
using that IMF. We have already heard during this week that nobody has
found overwhelming evidence for variations in the IMF. But as absence
of evidence is not evidence of absence can we conclude that the IMF is
in fact universal? The reasons for suspecting that it might not be are
that physical considerations seem to tell us that the IMF ought to
depend on metallicity, gravity and interstellar medium (ISM) pressure,
as examples. What I am about to say is detailed in a poster by Casuso
(these proceedings, p.~000). In that work we were, using very simple
assumptions, able to make an initial uniform spectrum of cloud core
masses which is brought about by considering motion within a closed
box, e.g., a spiral arm, and letting the system evolve. Using an
algorithm in which the tendency of clouds in collision to merge or to
fission is determined by their masses and their mutual velocities, we
find results which agree well with what is measured locally in the
ISM, but are also stable over time. Our model uses an arbitrary
initial mass distribution of the clouds, and the output results depend
only slightly on the input parameters, and very little on epoch after
an initial relaxation time of order a few Myr. This model is
metallicity independent and virtually independent of dynamics. Its
simplicity and robustness may enable us to gain some insight into the
possible existence of a universal IMF.

\medskip

{\sc Balcells:} The confirmation that the color distribution of
galaxies is bi-modal has far-reaching consequences for our
understanding of galaxy evolution.  It has provided a simple, but
inspiring, unifying theme for galaxy evolution.  Galaxies are born and
grow in the \textit{blue cloud}, a sort of nursery for galaxies.
Eventually, galaxies come of age, and migrate to the \textit{red
sequence}, a sort of retirement paradise, where massive galaxies live
forever, growing only moderately by aggregation.

Irrespective of how much of this picture will eventually be shown to
correspond with reality, one obvious question arises: what physical
processes drive galaxies away from the blue cloud into the red
sequence?

One basic hypothesis is that SF stops as a result of gas exhaustion,
due to the ongoing SF.  A second hypothesis is that gas may get
consumed, removed and/or heated to virial temperatures during mergers.
A third hypothesis, developed by Dekel and collaborators, exploits the
presence of the most massive galaxies in the red sequence: the gas
accreting onto dark matter halos along the filaments of the cosmic web
remains cool when falling onto low mass halos, while it shock-heats to
virial temperatures in galaxies with masses above about
$\log(M/M_\odot) > 11.5$.  This model amounts to a phase transition in
the external interactions of galaxies.  A final hypothesis would be
that, as the potential of the galaxy deepens as a result of SF,
eventually a phase transition occurs to the galaxy's ISM, and a new
equilibrium, or quasi-equilibrium, is found at higher temperatures.

What makes the red sequence interesting is that it presents us with a
lot of physics to analyze, in a territory where theorists and
observers can contribute, compete, and, hopefully, lead us to a better
understanding of the workings of galaxies.

\medskip

\begin{figure}[!ht]

\includegraphics[scale=0.65]{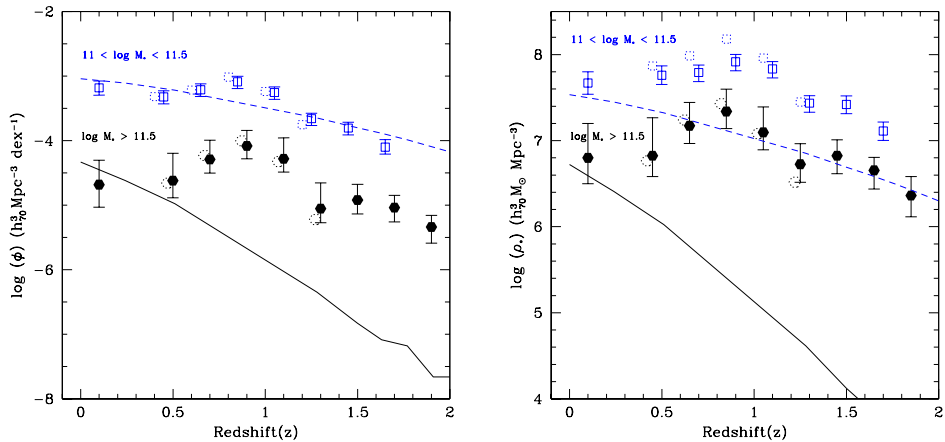}
\caption{A comparison between Palomar NIR survey data (Conselice et
al. 2007), and models from the Millennium simulation (e.g., De Lucia
et al. 2006).  The dashed line shows the predicted evolution in the
number and stellar mass densities for $10^{11}\,M_{\odot}\,< M^{*} <$
$10^{11.5}\,M_{\odot}\,$ systems, while the solid line shows the same
predicts for galaxies with stellar masses $M^{*} >$
$10^{11.5}\,M_{\odot}\,$.  As can be seen, for the most part the
simulated massive galaxies do not assemble quickly enough to match the
observations.}
\end{figure}

{\sc Conselice:} Figure~1 shows the mass and number densities for
massive galaxies as a function of redshift. As can be seen, the
densities for the most massive galaxies with $\log M > 11.5$ are
nearly all in place, statistically, at $z<2$.  The two solid lines
show the Millennium simulation results for the same quantities. As can
be seen, the Millennium results underpredict the number of massive
galaxies by up to two orders of magnitude.  Although there is good
agreement between the data and the models at $z = 0$, there is a large
disagreement at higher redshifts, which continues to grow. This shows
that the processes responsible for the assembly of massive galaxies in
the Millennium simulation, which are generally mergers, occur much
later in the simulation than for real galaxies.  The formation of
massive galaxies is therefore occurring much earlier than
predicted. It is also observed that galaxy mergers are common at $z >
2$, and it is likely that galaxy merging is not occurring early enough
in the simulations to produce the number of distant massive galaxies.

\medskip

{\sc Silk:} I wish to address the limitations of numerical simulations
of galaxy formation and evolution. The problem is one of inadequate
resolution. If one wishes to address SF one must go to very small
scales and an enormous dynamical range - beyond reach currently. New
rules are needed for sub-grid physics. Our current knowledge is based
on observational phenomenology. However, detailed observations made
locally are too complicated to implement, so assumptions must be
made. These are then fine-tuned as more detailed knowledge becomes
available, but it is a fundamental question as to how robust the
predictions are. This is a general statement, but here are a few
examples:

\begin{itemize}

\item There are three main problems with cold dark matter theory
  (CDM). These are the frequency of dwarf galaxies, of massive
  galaxies, and the ubiquity of halo cusps. The solutions to these
  problems may be tracers of early SF. Why is it that the cusps in
  galaxy centers cannot be erased by, e.g., a bar? Models look at the
  effect of the bar on its surroundings, but a bar is accompanied by
  inflow, and by continuous re-formation. So one must study the
  frequencies of dwarf and massive galaxies in different environments
  and metallicities, as well as the cores of halos. Any correlations
  obtained may lead to potential insights on fossil SF.

\item Disks formed stars inefficiently but massive galaxies formed
  stars very efficiently. These insights are due to the derivation of
  specific SF rates (SFRs) from spectrophotometric modeling, and from
  [$\alpha$/Fe] abundance ratio studies. One resolution may lie in
  invoking two modes of SF, with the inefficient disk mode being
  associated with supernova feedback and the efficient massive
  spheroid mode being at least partly regulated by AGN outflows.

\item The first ionizing sources at high redshift may have been
  massive stars, possibly with a top-heavy IMF, or mini-quasars. Are
  intermediate mass black holes required to account for re-ionization
  if the IMF is not top-heavy, and if so, what are their signatures?

\end{itemize}

My bottom line is that one should not believe current models,
especially as far as their predictive power is concerned.

\medskip

{\sc Kroupa:} I'd like to bring up two additional problems which may
well be associated with a currently wrong view of structure formation
and the build-up of stellar content of galaxies:

\begin{enumerate}

\item Joe (Silk) already mentioned that CDM theory predicts the wrong
number of dark matter dominated satellites of $L^*$ galaxies. In this
context I'd like to emphasize this problem further: we know that when
late-type galaxies interact, tidal arms take away the angular momentum
and energy.  When these are gas rich, galaxy-sized objects are often
observed to form in the arms; sometimes a dozen star-forming dIrr-type
knots are evident per interacting galaxy pair. Theoretical work shows
the number of self-gravitating knots to scale with the gas
fraction. Okazaki \& Taniguchi (2000) apply a standard hierarchical
structure formation prescription for the build-up of galactic mass,
and show that if only 1--2 long-lived tidal-dwarf galaxies (TDGs)
survive per encounter involving late-type galaxies then {\it all} dE
galaxies can be accounted for. In particular, their work shows that
the observed morphology-density relations for both dwarf and giant
galaxies in the field, groups of galaxies, and clusters of galaxies
are readily explained. Okazaki \& Taniguchi made very conservative
assumptions on the number of TDGs formed, and the actual number of
TDGs is likely to be much higher because the early galaxies had a
larger fraction of their mass in gas.

My own work (Kroupa 1997; Metz \& Kroupa 2007) shows that the
interpretation that the Milky Way dSph satellites are dark matter
dominated is probably wrong and that they are most probably ancient
TDGs. Their spatial distribution about the Milky Way supports this
conclusion (Metz, Kroupa, \& Jerjen 2007).

Taken together, this would suggest a major logical problem with CDM
theory because there would not be any room in the observed cosmos for
any type of dark matter dominated satellite galaxy; CDM theory would
predict the formation of more than enough TDGs to explain all known
dwarf galaxies. The problem is that these TDGs cannot be explained away
-- they are an inherent consequence of fundamental conservation laws.

\item In cosmological applications the stellar IMF is often taken to
be of an invariant Salpeter form. Sometimes top-heavy forms are
considered.  Recently it has emerged, though, that the stellar IMF of
a whole galaxy, the ``integrated galaxial IMF'' (IGIMF) is a result of
adding up all the stellar IMFs in all the star clusters that are
forming. The IGIMF turns out to be steeper than the invariant stellar
IMF and to depend on the SFR of a galaxy. Low-mass galaxies have
steeper IGIMFs than massive galaxies and also a larger variation of
the IGIMF. This has important implications for extragalactic
astrophysics, as, e.g., the mass-metallicity relation of galaxies
would be a natural result of such a variation of the IGIMF (K\"oppen,
Weidner \& Kroupa 2007). Also, the H$\alpha$-SFR calibration would be
wrong for low-mass galaxies; the observed H$\alpha$ fluxes would imply
significantly larger true SFRs than currently thought
(Pflamm-Altenburg, Weidner \& Kroupa 2007). Clearly, this needs to be
taken into account in future stellar population studies.

\end{enumerate}

I would now like to propose the following six points to discuss with
the audience:

\begin{enumerate}

\item What is the mass function of pre-stellar gas clouds,

\item Can we distinguish between a  top-heavy IMF at high redshifts,
  and mini-quasars as the first ionizing sources,

\item Is there a galaxy-wide IGIMF, along the lines of the
  description given by Weidner \& Kroupa (2005) -- we know the
  Salpeter description does not work,

\item Are there differences in the specific SFRs and in the
[$\alpha$/Fe] relative abundances between disks and massive galaxies,

\item Do merging rates yield a wrong prediction for massive galaxies,
and how does this relate to the red sequence,

\item Are apparently dark matter-dominated dwarf galaxies really TDGs,
or is theory trying to explain something that isn't there?

\end{enumerate}

Can I invite comments from the audience please?

\medskip

{\sc Pagel:} According to Kashlinsky et al. (2007), a small
contribution to the extragalactic background light could come from
supermassive population III objects for which they have measured an
angular correlation function resembling that of galaxies.

\medskip

{\sc Kroupa:} And we should not just consider the dark matter
paradigm, but also different gravitational laws to study where our
understanding breaks down.

\medskip

{\sc Alves:} If the characteristic mass function of the IMF is the
critical Bonnor-Ebert mass (cf. Alves, Lombardi \& Lada 2007), then I
would naively expect, at least, a bottom-deficient IMF for the first
stars, as $M_{\rm BE}\propto c^2/p^{0.5}$. 

\medskip

{\sc Silk:} The back-of-the-envelope calculation, which uses a sound
speed approximation to H$_2$ cooling, indeed gives about
1000\,$M_\odot$ for the mass of the first stars to form in a
metal-free environment. High dynamic range simulations of the first
stars initially confirmed the naive predictions for the {\it minimum}
mass. More recent simulations, which are now able to study more than
one realization of collapse over a wide dynamic range, find a
dispersion in the central temperature and density gradients that
controls the accretion and angular momentum transfer rates. A range in
first star mass is found, down to as low as 10\,$M_\odot$, and up to
100\,$M_\odot$. These stars are in the normal stellar mass range.

\medskip

{\sc Krumholz:} Theoretical models and simulations seem to suggest
that the Arches cluster should have a top-heavy IMF, but observations
have shown that the Arches IMF is basically the same as everywhere
else in the Galaxy. Given the failure of our models for the Arches, we
should be wary of applying theoretical models that predict varying
IMFs to the high redshift universe.

\medskip

{\sc Dom\'\i nguez-Tenreiro:} I would like to comment on the issue of
the top-heavy IMF versus mini-quasars as the first ionizing
source. Cosmic flows can be described on a mathematical level by the
Burgers equation, a generic equation in the framework of non-linear
physics. This equation predicts that flows unavoidably develop
singularities (i.e., black holes in the current context) under generic
conditions. So, it is very likely that mini-quasars appear very early
in the evolution of the Universe, and they would act as sources of the
first ionization. Of course, this does not exclude a top-heavy IMF at
high redshift, too.

\medskip

{\sc Peletier:} I would like to reiterate an up to now unsolved
problem: the low Ca{\sc ii} IR triplet values in giant
ellipticals. This was published by Saglia et al.  (2002), Cenarro et
al. (2003) and Falc\'on-Barroso et al. (2003). Assuming a normal
Salpeter or Kroupa IMF, an average old age, and a high metallicity,
necessary to fit absorption lines in the optical, the derived Ca
abundance is lower than solar, unless the IMF is top-heavy at the
low-metallicity end. Below solar [Ca/Fe] ratios are in principle
unattractive, since Ca is an alpha-element, and alpha-elements are
generally over-abundant with respect to Fe in giant ellipticals. The
observations are beyond question, and the community is still waiting
for a satisfactory solution for a line which is relatively
well-understood.

\medskip

{\sc Shlosman:} I would like to respond to Joe's (Silk) apparent
criticism of numerical simulations of dark matter cusp destruction by
stellar bars. I fully agree that results of simulations should always
be taken with a healthy dose of skepticism, {\it unless} they are
supported by compelling theoretical arguments and common sense. In the
particular case of the dark matter cusps, all numerical simulations
with live bars (including our still unpublished results of
$N\sim10^{10}$ particles by Dubinski, Shlosman \& Berentzen) show that
the cusps do {\it not} dissolve (except in simulations by
Holley-Bockelmann). Moreover, there are no compelling theoretical
arguments as to why this should occur. The models now encompass pure
collisionless cases as well as gaseous disks.

\medskip

{\sc Silk:} Proto-galactic bars in a gas-rich environment have not yet
been adequately studied. I do not think the dust has settled yet on
the issue of cusp softening.

\medskip

{\sc Vazdekis:} [$\alpha$/Fe] overabundance ratios are usually assumed
to be a result of highly efficient SF. But if the prediction by, e.g.,
Schneider et al. (2003) of a top-heavy IMF for metallicities lower
than $10^{-5}$ is true, these overabundance ratios can be explained in
part without invoking such high SF efficiency.  Furthermore, a
combination of the two scenarios might be appropriate for explaining
the observed overabundance ratios.

\medskip

{\sc Kroupa:} But if one would have a top-heavy IMF, could one make a
higher [$\alpha$/Fe] without the need for a higher SF efficiency?

\medskip

{\sc Silk:} As you must also produce the Fe, you need type Ia
supernovae, and so a top-heavy IMF will not be enough.

\medskip

{\sc Pipino:} I would like to make two comments on points that were
previously raised. Firstly, on Vazdekis' quest for population III
stars, we showed (Matteucci \& Pipino 2005) that if you take
population III stars into account as a real first (and single)
generation of stars, you wouldn't notice their effect after
$10^{11}\,M_\odot$ of stars have been created with a standard Salpeter
IMF, over a period of at least half a Gyr. Secondly, on the
Ca depletion in ellipticals: it is a false problem because Ca is also
produced by type I supernovae in a non-negligible way (up to half of
the total mass can be synthesized by supernovae Ia for a standard
IMF), as is Si, therefore a ratio of [Ca/Mg]$\leq0$ is naturally
predicted by chemical evolution models (as described in Pipino \&
Matteucci 2004).

\medskip

{\sc Peletier:} This paper still claims that Ca, as an alpha-element,
is overabundant with respect to Fe, while in these ellipticals [Ca/Fe]
lower than solar seems to be needed. I agree that we have to look for
the solution also in the direction of the chemical enrichment models.

\medskip


{\sc Hensler:} As one learns from changes in element production in
massive stars by refinements of the stellar models over the last
years, a word of caution should be emphasized not to rely too much on
abundances and their relations and take them too seriously as tracers,
e.g., of changes in the IMF and SF episodes.  So I agree that we
should be more careful with interpretations of models until we have
made sure that all the relevant physical processes are properly
included, and until their influences on galaxies within the complex
network of matter cycles are intensively explored.

\medskip

{\sc Hammer:} I would just like to question the assumption that
considering their specific SFRs, disk SF is not efficient. This is
certainly true for local disks but what about the period six to eight
Gyr ago?  Some 15\% of the $M^*$ galaxy population is made with
luminous infrared galaxies (LIRGs) which can double their mass in a
few times $10^8$\,yr. And almost half of them look to be disks, some
even with a normal-looking rotation curve.

\medskip

{\sc Balcells:} Maybe a way of addressing the question is to study
whether LIRGs can be understood in terms of quiescent or burst-mode
SF, or whether there is some continuum between the extremes.

\medskip

{\sc Hammer:} Well, spectroscopy shows that in general LIRGs are the
same as, or very similar to, spirals, which is a purely observational
result.

\medskip

{\sc Krumholz:} I object, as I did in the first panel discussion and
on other occasions, to the use of the term efficiency. Unless we have
ironclad evidence that high-redshift SF does {\it not} follow the same
rules as local SF, e.g., the universal IMF and the Schmidt-Kennicutt
and Gao-Solomon correlations, we should not hypothesize the existence
of additional ``modes'' of SF. It is significant that local LIRGs and
ULIRGs follow the same correlations as normal galaxies, and, from what
we can tell, high-redshift ones do too.

\medskip

{\sc Hammer:} How do you know this ? There may be no different physics
at high $z$ but certainly different conditions (gas, merging
occurrence, etc...).

\medskip

{\sc Krumholz:} I repeat, the basic point is that we should only
introduce new physics if we really have to.

\medskip

{\sc Silk:} So ULIRGs lie on the Kennicutt law?

\medskip

{\sc Krumholz:} Yes indeed, local ULIRGs do lie on both the Kennicutt
law and the Gao-Solomon law.

\medskip

{\sc Balcells:} From the point of view of a galaxy observer, there are
indeed different modes, for instance, ellipticals have an
[$\alpha$/Fe] ratio that is not found in disks, which implies the
existence of a burst versus a quiescent mode. So I think we can talk
about two modes.  The two modes must reflect different physical
conditions -- though not necessarily new physical processes.

\medskip

{\sc Stringer:} Just a short comment: do different modes of SF
necessarily need to correspond to different IMFs?

\medskip

{\sc Beckman:} Not necessarily.

\medskip

{\sc Silk:} Could be the IMF, or it could be due to AGN.

\medskip

{\sc Kroupa:} This is a question of interpretation -- some panel
members would suggest no. Personally I think that under extreme
star-forming rates ($>10^2\,M_\odot\,{\rm yr}^{-1}$) a top-heavy IMF
does emerge. One indication of this is that the metallicity
distribution of the Milky Way bulge and of the M31 bulge are easily
reproduced with a top-heavy IMF, while a universal IMF does not match
the observations (Ballero, Kroupa, \& Matteucci 2007). In Bonn we are
working on constraining the IMF for extreme starbursts. It is too
early to state definite results, but a top-heavy IMF appears to be
necessary.

\medskip

{\sc Kroupa:} I need to start wrapping up, and was wondering whether
anyone wants to comment on our fifth point, on merging rates and the
red sequence.

\medskip

{\sc Conselice:} The merger rate or fraction is a very difficult
parameter to determine, and observations presently disagree with
theory. Much more work needs to be done before we can make statements
on this.

\medskip

{\sc Trujillo:} The small sizes of galaxies at a redshift of around
1.5 are not in contradiction with current $N$-body and semi-analytical
simulations. A couple of major mergers will locate these objects
within the local stellar mass-size relation (Boylan-Kolchin, Ma \&
Quataert 2006; Khochfar \& Silk 2006).

\medskip

{\sc Conselice:} The smaller sizes of these galaxies also demonstrate
that merging is occurring, but how the observed merger rate, itself,
agrees or not with the models still needs to be determined as there is
no clear cut answer at the moment.

\medskip

{\sc Kroupa:} Let's hear a final comment from the audience before we
turn back to the panel?

\medskip

{\sc Hammer:} There is a lot of emphasis on the reproduction of the
merger rate by simulations, but there is also a question to
observers. In 1999 we made the first observation of pairs and of the
merger rate up to a redshift of unity in the framework of the
Canada-France Redshift Survey (CFRS). Since that time a very large
number of papers has appeared on the subject, with different
assumptions and methods (evolution-corrected or not, using blue light,
red light, etc.). The overall result, also on the observers' side, is
real confusion. I suggest that we should resolve this confusion before
going to different simulations which also have various
predictions. Otherwise, this ``problem'' will never end!

\medskip

{\sc Balcells:} The observational results can also usually be changed
with sample selection.  In work from our group, $B$-band vs $K$-band
selection yields different $z$-evolution of the merger fraction, even
when $I$-band asymmetries are used for both diagnostics.  Pixellation
and cosmological dimming need to be carefully calibrated when using
asymmetries or pair fractions to infer the $z$-evolution of the merger
fractions.  Going beyond $z\sim1$ becomes especially tricky as
CCD-based imaging surveys start to sample the rest-frame UV continuum.
Given these difficulties, disagreement between different observers is
almost unavoidable.

\medskip

{\sc Conselice:} Merger fractions can be calculated in various ways
and need to be determined quantitatively before comparing them with
observations.

\medskip

{\sc Silk:} On the one hand the merger rate does not agree well with
the models. But on the other, I am sure that modelers are very clever
and so will come up with new models which will fit much better. But
the question is whether we will learn any new physics from it?

\medskip

{\sc Kroupa:} We have run out of time, so before finalizing this panel
discussion I would like to ask John (Beckman) to make a final
statement.

\medskip

{\sc Beckman:} It is refreshing to see so much discord about the
IMF. There are good reasons for uncertainty about the question of
whether under extreme star-forming conditions or under conditions of
very low metallicity the IMF is non-standard. These have been
mentioned in the discussion and are as always a question of how to
interpret the observations, notably of element abundance ratios. The
question of whether population III occurred or not is closely linked
with this and is also up for grabs for the same sort of reasons.  It
is also refreshing to find that there are no clear answers to the
problem of the lack of cuspiness in galaxy centers, or indeed to the
question of why the semi-analytical models, with mergers at their
heart, give such a relatively poor account of the evidence from
stellar populations of the way and the rates at which galaxies have
evolved. All this bodes well for the next generation of researchers.

\end{document}